\documentclass[prl,aps,twocolumn,superscriptaddress]{revtex4-2}
\usepackage{graphicx,color}
\usepackage{amsthm}
\usepackage{amsfonts}
\usepackage{algorithmic}
\usepackage{enumerate}
\usepackage{latexsym}
\usepackage{amsmath}
\usepackage{amssymb}
\usepackage{subfig}
\usepackage[colorlinks=true,citecolor=blue,linkcolor=blue]{hyperref}

\usepackage{dcolumn}
\usepackage{bm}

\usepackage[justification=raggedright]{caption}
\usepackage[T1]{fontenc}
\usepackage{mathptmx}

\DeclareMathOperator{\Tr}{Tr}

\begin{document}

\title{Parameters dependent superconducting transition temperature in high temperature superconductors}

\author{Runyu Ma}
\affiliation{School of Physics and Astronomy, Beijing Normal University, Beijing 100875, China\\}
\affiliation{Key Laboratory of Multiscale Spin Physics(Ministry of Education), Beijing Normal University, Beijing 100875, China\\}
\author{Tianxing Ma}
\email{txma@bnu.edu.cn}
\affiliation{School of Physics and Astronomy, Beijing Normal University, Beijing 100875, China\\}
\affiliation{Key Laboratory of Multiscale Spin Physics(Ministry of Education), Beijing Normal University, Beijing 100875, China\\}

\author{Congjun Wu}
\email{wucongjun@westlake.edu.cn}
\affiliation{New Cornerstone Science Laboratory, Department of Physics, School of Science, Westlake University, Hangzhou 310024, Zhejiang, China\\}
\affiliation{Institute for Theoretical Sciences, Westlake University, Hangzhou 310024, Zhejiang, China\\}
\affiliation{Key Laboratory for Quantum Materials of Zhejiang Province, School of Science, Westlake University, Hangzhou 310024, Zhejiang, China\\}
\affiliation{Institute of Natural Sciences, Westlake Institute for Advanced Study, Hangzhou 310024, Zhejiang, China\\}

\begin{abstract} 
Understanding the evolution of the superconducting transition temperature in relation to doping and interaction
strengths is one of the most challenging problems in high temperature superconductivity. By refining determinant quantum Monte Carlo algorithm,
we characterize the parameter dependence of the superconducting transition temperature within a bilayer Hubbard model, which is sign-problem-free at arbitrary filling. A striking feature of this model is its similarities to the bilayer nickelate-based superconductor $\mathrm{La}_{3}\mathrm{Ni}_{2}\mathrm{O}_{7}$, where superconductivity emerge from the bilayer $\mathrm{Ni}\mathrm{O}_{2}$ planes.
We find that interlayer spin-exchange $J$ is critical to interlayer pairing, ant that on-site interaction $U$ contributes negatively to superconductivity at low doping levels but positively to it at high doping levels.
Our findings identify the key parameter dependent superconducting transition temperature in nickelate-based superconductors and provide a new understanding of the high temperature superconductivity.
\end{abstract}

\maketitle

\noindent
\underline{\it Introduction}~~
Since the discovery of unconventional superconductivity in doped cuprates\cite{PhysRevLett.58.908,Bednorz1986}, understanding the parameter dependence of superconducting transition temperatures $T_{c}$
has been one of the most crucial and challenging task in current condensed matter physics and material science. Experimentally, researchers are continually seeking out new unconventional superconducting systems, hoping to discover ones with even higher superconducting transition temperatures.
After few decades of intensive studies, a series of superconducting material have been developed, including multilayer copper-based superconductors\cite{Luo2023,Yamamoto2015,Chen2010},
iron-based superconductors\cite{doi:10.1021/ja800073m,Yi2017,doi:10.1126/science.aab0103,doi:10.1080/00018732.2010.513480,Paglione2010,RevModPhys.83.1589},
and the most recent discovery of nickelate-based superconductors\cite{Li2019,Sun2023}.
Theoretically, characterizing the parameter dependencies of superconducting transition temperatures has proved to be
especially difficult as not only an effective microscopic model is stilling locking or uncertain, but also,
its dependence have been shown to be exceptionally sensitive to small changes in the model terms and parameters, neither temperature nor interaction, or doping.
The Hubbard model or its extension maybe one platform to investigate this problem. However, the relevant model parameters are in the most difficult regime where most approaches struggle\cite{science.adh7691,science.aam7127}, which has remained the central model studied by theorists. 

Therefore, developing an effective model that can be solved exactly providing the only opportunity to explore the underlying mechanism of superconductivity and the factors that affect superconducting transition temperature\cite{doi:10.1146/annurev-conmatphys-090921-033948, PhysRevX.5.041041, PhysRevX.11.011058}.
After intensive efforts for decades, a possible way to win this great challenge has emerged: applying the determinant quantum Monte Carlo (DQMC) method to
a sign-problem-free microscopic model which analogous to real materials.
A proposed sign-problem-free model is a bilayer system where the interactions have time-reversal symmetry after Hubbard-Stratonovich
transition\cite{PhysRevLett.91.186402}, in which contains
on-site Hubbard interaction and interlayer spin-exchange\cite{PhysRevB.106.054510, PhysRevB.107.214509}.
However, this sign-problem-free model deviates from existing superconducting systems in some details, until the realization of superconductivity in nickelate-based material\cite{Li2019,PhysRevLett.125.027001,PhysRevLett.125.147003}.

Superconductivity in nickelate-based materials is first observed in the family of infinite-layer nickelate thin-film materials
$\mathrm{A}_{1-x}\mathrm{B}_{x}\mathrm{Ni}\mathrm{O}_{2}$ ($\mathrm{A} = \mathrm{La},\mathrm{Nd},\mathrm{Pr}$, $\mathrm{B}=\mathrm{Sr},\mathrm{Ca}$)\cite{Li2019,PhysRevMaterials.4.121801,PhysRevLett.125.147003,doi:10.1126/sciadv.abl9927}.
Sooner, signatures of superconductivity in $\mathrm{La}_{3}\mathrm{Ni}_{2}\mathrm{O}_{7}$ (LNO) are observed\cite{Sun2023}, and its transition
temperature $T_{c}$ reaches 80K under high pressure. Besides its superconductivity, the magnetism and other properties of LNO
are also investigated\cite{Liu2022,zhang2023hightemperature}.
From a theoretical point of view, the electronic properties of LNO are determined by the bilayer $\mathrm{Ni}\mathrm{O}_2$ planes,
considering $d_{x^2-y^2}$ and $d_{z^2}$ orbitals, which can capture most of the low-energy electronic structures\cite{gu2023effective, PhysRevLett.131.126001},
and these two orbitals are essentially associated with the superconductivity in LNO.
To explain the pairing mechanism of LNO, various interaction terms are proposed to construct an effective model of LNO and describe its superconductivity.
A possible mechanism is interlayer $s$-wave pairing induced by the interlayer spin-exchange of either $d_{x^2-y^2}$ or $d_{z^2}$ orbitals
\cite{yang2023minimal, lu2023interlayer, kaneko2023pair}. The interlayer spin-exchange of $d_{z^2}$ originates from strong interlayer hopping, and the interlayer spin-exchange of $d_{x^{2}-y^{2}}$
is passed from $d_{z^2}$ due to Hund's rule. Under the large Hund coupling, $d_{x^2-y^2}$ spins are forced to align with to $d_{z^2}$, and integrated out $d_{z^2}$ degrees of freedom resulting in an effective interlayer spin-exchange
in $d_{x^2-y^2}$ orbital\cite{lu2023interlayer,qu2023bilayer}. Then a reduced Hamiltonian can be derived, which is a bilayer single orbital model and contains
large interlayer spin-exchange.
Besides, in some other works, superconductivity in LNO is attributed to its similarity with doped cuprates\cite{PhysRevB.108.L201121, jiang2023high, fan2023superconductivity}.

The reduced low-energy effective model of nickelate-based superconductors is quite similar to
the sign-problem-free microscopic model, which provide us a rare opportunity to study the parameters dependent $T_c$ numerically exactly based on
a realistic  model closed related with real materials.
In this work, we make an initial step on determining
the parameters dependent superconducting transition temperature $T_c$  numerically exactly.
We characterized $T_c$ by using the universal-jump relation of
superfluid density at a large region of
electron filling $\langle n \rangle$ and coupling strength $U$. Interestingly, we find that $U$ benefits to superconductivity
at high doping levels and does harm at low doping levels, and we also find that $t_{\bot}$ does harm to superconductivity.
Finally, we demonstrate the regular part of resistivity $R^{reg}$, which is helpful for us to deduce the Drude weight and study the strange-metal behavior.
These results demonstrate the potential of this bilayer sign-free model in studies of LNO and superconductivity, and
may also help settle the problem of the pairing mechanism of LNO.

\noindent
\underline{\it Model \& Method}~~
The sign-problem-free microscopic model we used can be written as
\begin{equation}
    \begin{aligned}
        H = & -t\sum_{\langle i,j \rangle} \left(\psi^{\dagger}_{i} \psi_{j} + H.c.\right)
        - \mu \sum_{i} n_{i} -t_{\bot}\sum_{i} \psi^{\dagger}_{i} \Gamma^{~4} \psi_{i} \\
        &  - \sum_{i,a=1\text{-}5} \frac{g_{a}}{2} \left( \psi^{\dagger}_{i} \Gamma^{~a} \psi_{i} \right)^2 -
        \sum_{i} \frac{g_{0}}{2}\left(\psi^{\dagger}_{i} \psi_{i} - 2\right)^{2}
    \end{aligned}
    \label{eq:1}
\end{equation}
where we define $\psi_{i} = \left(c_{i1\uparrow}, c_{i1\downarrow}, c_{i2\uparrow}, c_{i2\downarrow} \right)^{T}$, $n_{i} = \psi^{\dagger}_{i} \psi_{i}$ and $c_{il\sigma}$ is the annihilation operator with site $i$, layer $l$, spin $\sigma$.
The $\Gamma$ matrix we used here is
\begin{equation}
    \begin{aligned}
        \Gamma^{1\text{-}3} = \begin{pmatrix}
            ~\vec{\sigma}~ & ~0~ \\
            ~0~ & ~ -\vec{\sigma}~
        \end{pmatrix}
        ~
        \Gamma^{4} = \begin{pmatrix}
            ~0~ & ~ I~ \\
            ~I~ & ~ 0~
        \end{pmatrix}
        ~
        \Gamma^{5} = \begin{pmatrix}
            ~0~ & ~ iI~ \\
            ~-iI~ & ~ 0~
        \end{pmatrix}
    \end{aligned}
\end{equation}

\begin{figure}
    \centering
    \includegraphics[width=0.48\textwidth, trim=0 0 0 0, clip]{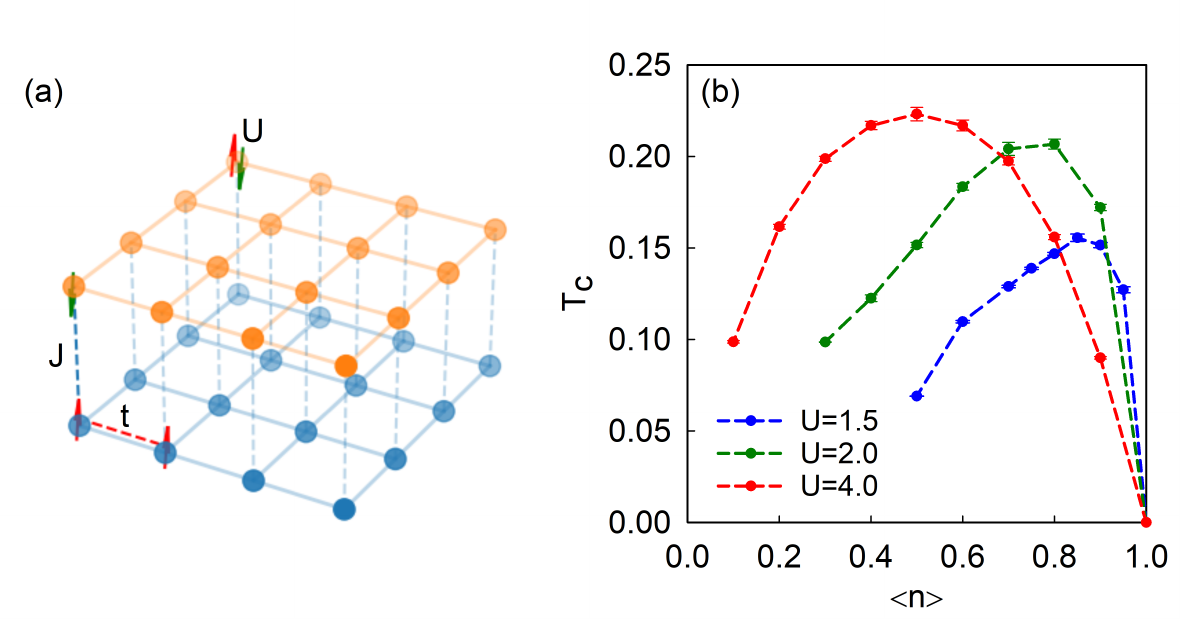}
    \caption{(a) A sketch of the bilayer model used in this work, periodic boundary condition is imposed in both directions.
    (b)Transition temperatures $T_c$ of different filling $\langle n \rangle$ from $U=1.5$ to $U=4.0$.}
    \label{fig:latt}
\end{figure}

For convenience, we rewrite the Hamiltonian Eq.~\ref{eq:1} in the following form
\begin{equation}
    \begin{aligned}
        H = & -t\sum_{\langle i,j \rangle l \sigma} \left(c^{\dagger}_{il\sigma} c_{jl\sigma} + H.c.\right) - t_{\bot}\sum_{i\sigma}\left( c^{\dagger}_{i1\sigma} c_{i2\sigma} + H.c. \right)   \\
        & - \mu \sum_{il\sigma} n_{il\sigma} + U \sum_{il} \left( n_{il\uparrow} - \frac{1}{2} \right) \left( n_{il\downarrow} - \frac{1}{2} \right)\\
        & + J_{z} \sum_{i} S^{z}_{i1} S^{z}_{i2} + \frac{J_{\bot}}{2} \sum_{i} \left( S^{+}_{i1} S^{-}_{i2} + H.c. \right)
    \end{aligned}
    \label{eq:3}
\end{equation}
where $S^{z}_{il} = \frac{1}{2}\left(n_{il\uparrow} - n_{il\downarrow}\right)$, $S^{+}_{il} = c^{\dagger}_{il\uparrow} c_{il\downarrow}$, and
$n_{il\sigma} = c^{\dagger}_{il\sigma} c_{il\sigma}$. $S^{-}_{il}$ is the hermit conjugate of $S^{+}_{il}$,
$U=g_{1} + g_{2} + g_{3} - g_{0}$, $J_{z} = 4g_{3}+4g_{0}$ and $J_{\bot} = 2g_{1} + 2g_{2} + 4g_{0}$.
Without specification, $g_{0}$, $g_{4}$ and $g_{5}$ is set to be $0$ and $g_{1}=g_{2}=g_{3}$, these make $J_{\bot}=J_{z}=4U/3$.
An illustration of our model is shown in Fig.~\ref{fig:latt}(a).
To some extent, this model is similar with the bilayer
nickelate model proposed in Ref\cite{lu2023interlayer, yang2023minimal} and mixed-
dimensional (mixD) model proposed in Ref\cite{schlömer2023superconductivity}, except that the spin-exchange in our work may be obviously large.

To investigate the properties of this model, we use DQMC algorithm to simulate the observables at finite temperature,
$\langle O \rangle = \Tr{e^{-\beta H} O} / \Tr{ e^{-\beta H}}$. In DQMC algorithm, the $e^{-\beta H}$ term will be
discretized into small pieces, $e^{-\beta H} = \prod_{M} e^{-d\tau H}$, where $M d\tau = \beta$. Then the interaction part of Hamiltonian will
be decoupled by Hubbard-Stratonovich (HS) transformation, $e^{-d\tau H} = e^{-d\tau H_{0}} \prod_{i,a=1\text{-}3} \sum_{l_{ia}} \gamma_{l_{ia}} e^{\eta_{l_{ia}} \Gamma^{a}} = \sum_{\mathbf{l}=(l_{1},l_{2},...)} \mathbf{B}(l)$.
After that, the observables
can be treated as a summation of a series of noninteracting systems. For more details of DQMC algorithm, please see Ref\cite{2008Assaad,PhysRevB.40.506}.

The sign problem occurs when the determinant $P(C)$ is not positive definite. The usually reweight process make
the observables become $\frac{\sum_{C} \mathbf{P}(C) \mathbf{O}(C) / \sum{\vert \mathbf{P}(C) \vert}}{ \sum{ \mathbf{P}(C)} / \sum{\vert \mathbf{P}(C) \vert}}$;
and define $S(C) =  \frac{\mathbf{P}(C)}{\vert \mathbf{P}(C) \vert}$, it becomes
$\frac{\sum_{C} \vert \mathbf{P}(C) \vert S(C) \mathbf{O}(C) / \sum{\vert \mathbf{P}(C) \vert}}{ \sum{ \vert \mathbf{P}(C) \vert S(C)} / \sum{\vert \mathbf{P}(C) \vert}}$.
Then common Monte Carlo process can be done for both denominator and numerator. However, when positive terms and negative terms of $P(C)$ are nearly cancelled,
the denominator will have a large statistic error compare to its value. The accuracy of simulations will be affected.
A special family of models can make sure that $P(C)$ is positive, these models have Kramers symmetric interaction operator $V$. The model used in this work
is one of this family. More details about these models can be seen at Refs\cite{PhysRevB.71.155115,PhysRevLett.91.186402}.

\noindent
\underline{\it Results \& Discussions}~~
As previous works\cite{PhysRevB.106.054510}, the superconducting order parameter can be defined as
\begin{equation}
    O_{SC} = \frac{1}{\sqrt{2}} \sum_{i} \left(c_{i1\uparrow} c_{i2\downarrow} - c_{i1\downarrow} c_{i2\uparrow} \right)
\end{equation}
To identify the superconducting transition temperature $T_{c}$, we analyze the superfluid
density $\rho_s$, and we briefly describe the calculation of $\rho_s$ below\cite{PhysRevLett.68.2830, PhysRevB.47.7995}.

First, we define the $x$ component of current density operator
\begin{equation}
    j_{x}\left(i, \tau\right) = e^{\tau H}\left[\sum_{l=1,2}
    \sum_{\sigma=\uparrow,\downarrow} -ic^{\dagger}_{i+\mathbf{x}l\sigma} c_{il\sigma} + ic^{\dagger}_{il\sigma}c_{i+\mathbf{x}l\sigma} \right]e^{-\tau H}
\end{equation}
and the corresponding responses can be expressed as follows,
\begin{equation}
    \Lambda_{xx}(\mathbf{q}) = \frac{1}{N_{site}} \sum_{i, j} \int^{\beta}_{0} d\tau e^{i\mathbf{q}\left( \mathbf{R}_{i} - \mathbf{R}_{j} \right)} \langle j_{x}\left(i,\tau \right) j_{x}\left(j,0\right) \rangle.
\end{equation}
The superfluid density is the difference of the limiting longitudinal and transverse responses of current-current correlations
\begin{equation}
    \rho_s = \frac{1}{4} \left( \Lambda^{L} - \Lambda^{T} \right)
\end{equation}
where
\begin{equation}
    \begin{aligned}
        &\Lambda^{L} = \lim_{q_{x} \rightarrow 0} \Lambda_{xx}\left(q_{x}, q_{y}=0\right)\\
        &\Lambda^{T} = \lim_{q_{y} \rightarrow 0} \Lambda_{xx}\left(q_{x}=0, q_{y}\right)
    \end{aligned}
\end{equation}

At the Kosterlitz-Thouless (KT) transition, the universal-jump relation holds\cite{PhysRevB.105.184502, PhysRevLett.39.1201},
\begin{equation}
    T_{c} = \frac{\pi}{2} \rho^{-}_{s}
\end{equation}
here $\rho^{-}_{s}$ means the value of superfluid density below the critical temperature $T_c$.
To identify $T_c$, we plot $\rho_s\left(T\right)$ and find its intercept with $2T/\pi$.

From the calculated  current-current correlations, one can deduce the regular part of optical resistivity readily
\begin{equation}
    R^{reg} = \pi T^2 \left[ \frac{1}{N_{site}} \sum_{i,j} \langle j_{x}\left(i, \beta / 2 \right) j_{x}\left(j, 0\right) \rangle \right]^{-1},
    \label{eq:10}
\end{equation}
and this relation is applicable at relative high temperatures\cite{doi:10.1126/science.aau7063, PhysRevB.107.L041103}.
Although this value may not be a faithful representation of the dc transport, $R^{reg}$ is helpful for further investigation like deducing Drude weight\cite{li2023hubbard}.

Besides, the superconducting susceptibility $\chi_{SC}\left(\beta \right)$ is defined as
\begin{equation}
    \chi_{SC}\left(\beta \right) = \frac{1}{N_{site}} \int^{\beta}_{0}d\tau \sum_{i,j} \langle e^{\tau H} O_{SC} e^{-\tau H} O_{SC} \rangle.
\end{equation}
For KT transition, the superconducting susceptibility $\chi_{SC}$ satisfy the scaling relation $\chi_{SC} \sim L^{1.75} f\left(L/\xi \right)$.
At the transition point, $\chi_{SC}L^{-1.75}$ of different system size will cross each other since the $\xi$ goes to infinity.

\begin{figure}
    \centering
    \includegraphics[width=0.48\textwidth]{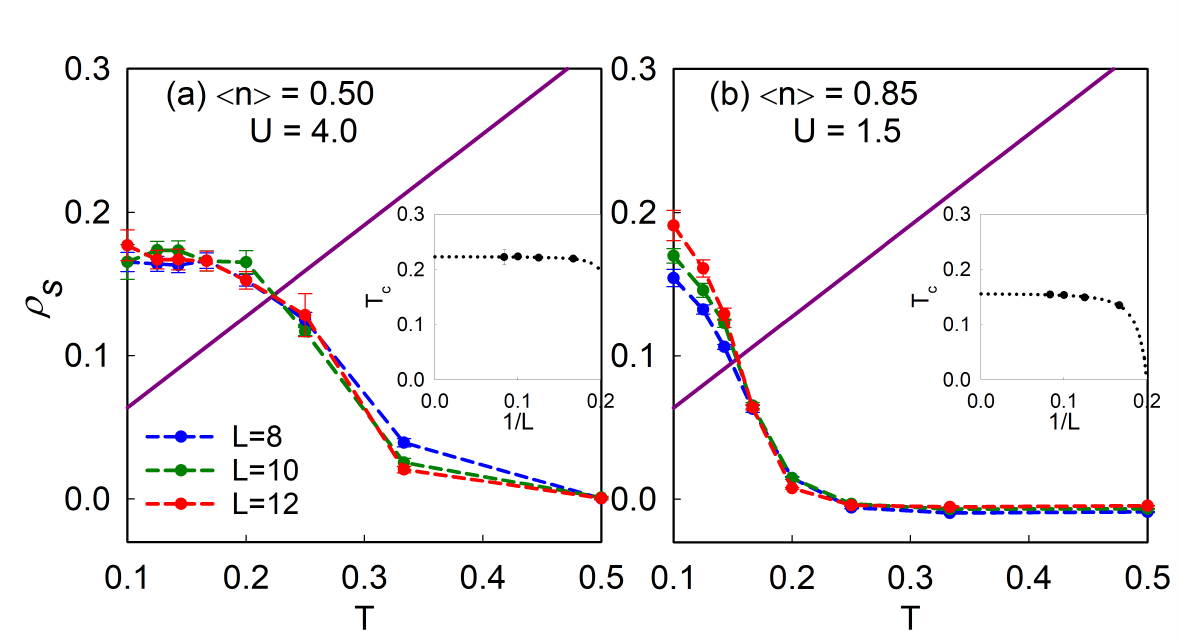}
    \caption{Superfluid density of (a) $U=4.0$, $\langle n \rangle = 0.50$ and (b) $U=1.5$, $\langle n \rangle = 0.85$.
    The straight line is $2T/\pi$. The superfluid density $\rho_s$ intersect the straight line at transition temperature $T_c$.
    Results are extrapolated to $L=\infty$ by fitting based on Eq.\ref{eq:fit}.
    }
    \label{fig:rhos}
\end{figure}

In Fig.~\ref{fig:rhos} we show the superfluid density $\rho_s$ of different system size $L$ and $J_{z}=J_{\bot}=4U/3$.
The size dependence of superfluid is weak in most case,
and we can fit the results based on the following equation
\begin{equation}
    T_{c}(L) = T_{c} + \frac{a}{\left(lnbL\right)^2}
    \label{eq:fit}
\end{equation}
In the following part of this work, we estimate the transition temperatures
by these extrapolations. 

\begin{figure}
    \centering
    \includegraphics[width=0.48\textwidth]{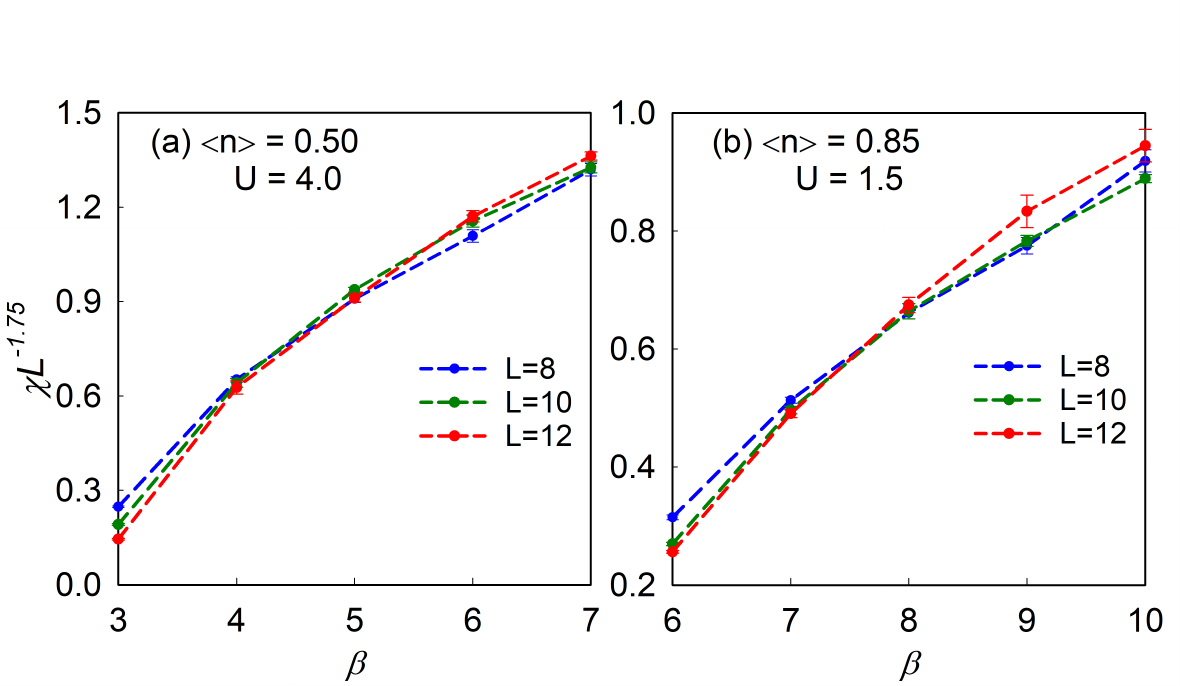}
    \caption{Pairing susceptibility of (a) $U=4.0$, $\langle n \rangle = 0.50$ and (b) $U=1.5$, $\langle n \rangle = 0.85$.
    Its value at difference system size $L$ intersects each other at transition temperature $T_{c}$.}
    \label{fig:chi1.75}
\end{figure}

To check the validity of these results obtained by $\rho_s$, we also examine the pairing susceptibility at different system size.
Results are shown in Fig.~\ref{fig:chi1.75}. The transition temperature $T_c$ is the crossing of different system size.
The crossing of $L=10$ and $L=12$
is near with the $T_{c}$ obtained by superfluid density $\rho_s$, and this confirms the validity of our results.
However,
the statistical error affects a lot, since curves of different $L$ are close to each other,
and the finite size effects also make it harder to extract the transition temperature.
For these reasons, we prefer to use superfluid density to estimate the transition
temperature.

Next we show $T_c$ at various electronic fillings.
Fig.~\ref{fig:latt}(b) shows transition temperatures $T_{c}$ from $U=1.5$ to $U=4.0$.
As $U$ increases, the optimal filling moves to $\langle n \rangle = 0.5$, close to the occupation of $d_{x^2-y^2}$ in single crystal LNO.
Recalling the definition of our model, the interlayer spin-exchange interaction also increases when $U$ becomes larger.
When $J$ is prevailing,
the superconducting dome shifts towards smaller electron filling\cite{chen2023orbitalselective} was found within the bilayer $t$-$J$ model,
and it was also suggested that an optimal doping at $\langle n \rangle = 0.5$\cite{lange2023pairing, lange2023feshbach} in the mixD+V model.
In our work where $U$ is present, the situation is more complicate.
Since $J$ will stabilize binding energy, the transition temperature $T_{c}$ increases monotonically with the increasing of $U$ is not surprising.
However, the situation at large $\langle n \rangle$ is rather interesting, where we can see that a non-monotonic dependence of $T_{c}$ on $U$ at small doping, namely,
$T_{c}$ increases at first, and further increasing $U$ will suppress superconductivity. Later we will show that this is originated from the competition of $U$ and $J$, and $U$ does harm to superconductivity at small doping. Since the complicated $U$ dependence, the optimal doping moves to $\langle n \rangle = 0.5$ as $U$ increases.

This diagram also demonstrate a possible strange-metal behavior owned by this model. Namely, $T_{c}$ is a quadratic function of doping.
This is found at the overdoped range of electron-doped copper oxide, and may have a common mechanism in unconventional superconductors\cite{Yuan2022}.
At $U=4.0$, $T_{c}$ have a clear quadratic form at both overdoped side and half-filled side($\langle n \rangle=0$).

\begin{figure}
    \centering
    \includegraphics[width=0.48\textwidth]{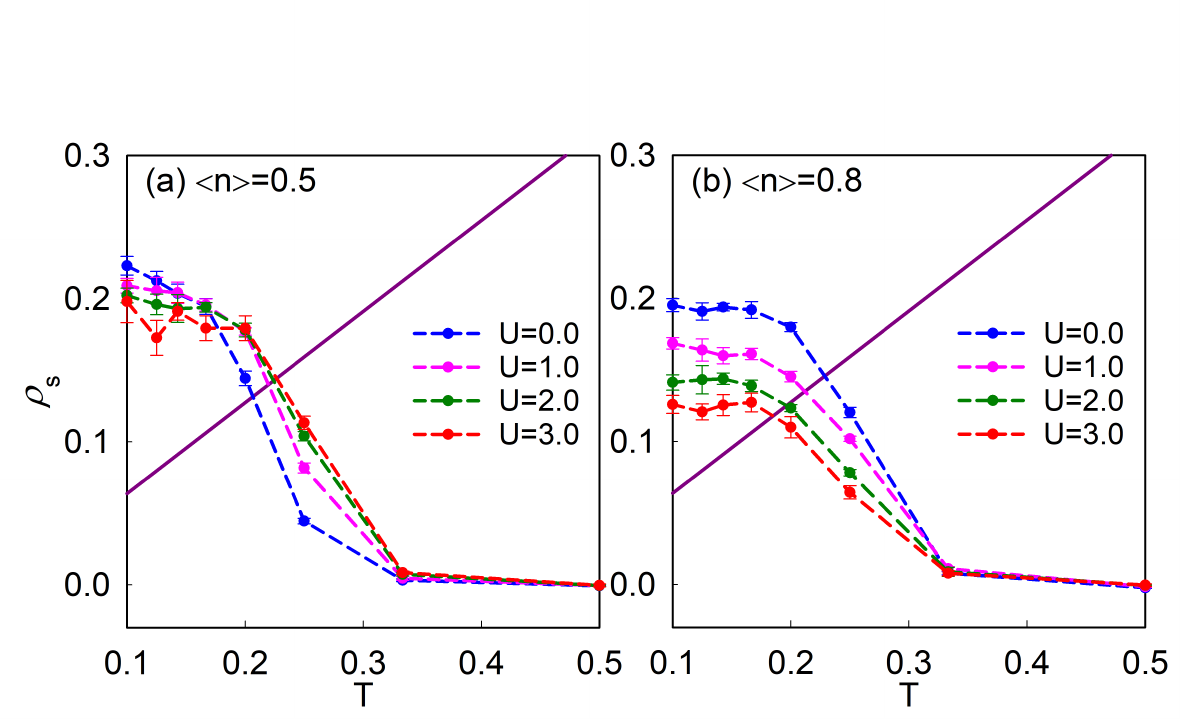}
    \caption{Superfluid density of different $U$ at $J_{z} = J_{\bot} = 4.0$ and $L=12$,
    at (a) $\langle n \rangle =0.5$ and (b) $\langle n \rangle =0.8$. Unlike other part of this work, $J = 4U/3$ does not hold in this diagram.}
    \label{fig:verU}
\end{figure}

Next we present the results with gradually suppressed $U$ and fixed $J$. To decrease $U$, we need to turn on the other three terms $g_{0}$, $g_{4}$ and $g_{5}$.
In the previous paragraph, we mentioned that dependence of $T_{c}$ on $U$ and the competition between $U$ and $J$.
In Fig.~\ref{fig:verU}, we demonstrate the behavior of superfluid density at
$L=12$ with different $U$ but fixed $J=4.0$. Here, we can witness the different dependence of $T_c$ on $U$.
At low doping level, $U$ could suppress $T_{c}$ when $J$ is fixed, and the superconductivity might be affected by other intertwined orders favored by $U$. This reveals an intrinsic
constraint on improving $T_{c}$, and also reveals that higher doping (near $0.5$) is more beneficial for interlayer superconductivity.
In LNO, since the interlayer spin-exchange originates
from interlayer hopping and $U$, the effects of $U$ could be very important.

\begin{figure}
    \centering
    \includegraphics[width=0.48\textwidth]{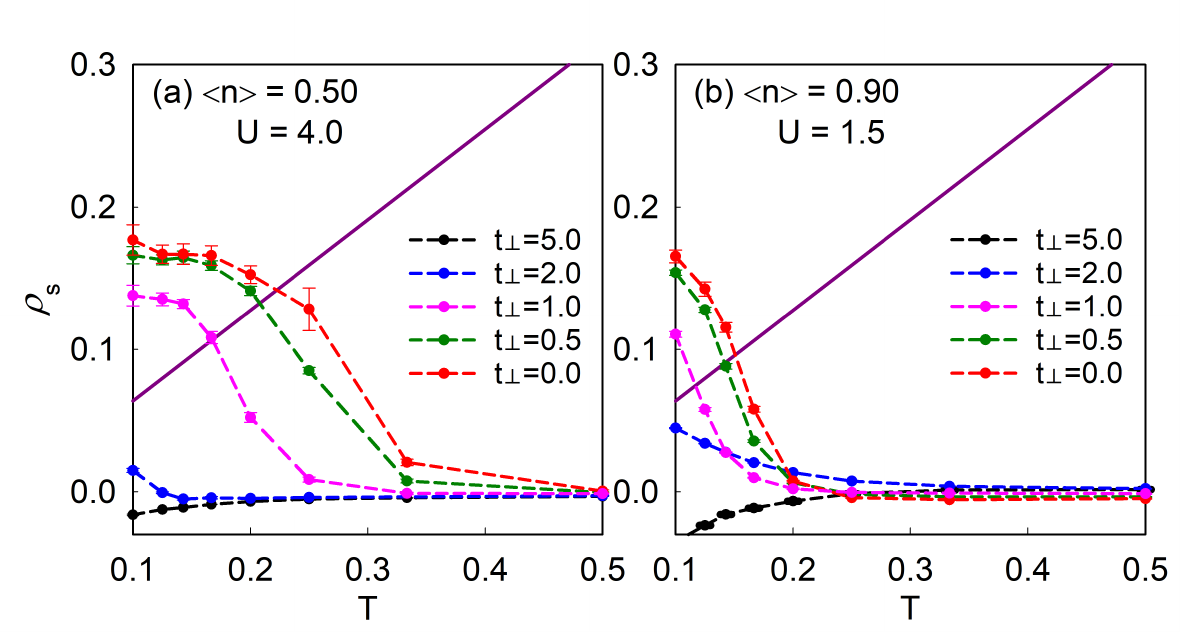}
    \caption{$\rho_s$ of (a) $U=4.0$, $\langle n \rangle=0.5$ and (b) $U=1.5$ $\langle n \rangle=0.9$. The superconductivity
    is suppressed when $t_{\bot}$ become larger.}
    \label{fig:tbot}
\end{figure}

Now we turn off $g_{0}$, $g_{4}$, $g_{5}$ terms and investigate the effects of $t_{\bot}$ term.
In LNO, $t_{\bot}$ shall modify the Fermi surface topology\cite{ryee2023critical}.
The interlayer hopping terms of $d_{x^{2}-y^{2}}$ and $d_{z^2}$ are different\cite{gu2023effective, PhysRevLett.131.126001},
and $d_{x^{2}-y^{2}}$ is much smaller than that of $d_{z^2}$.
As we can see in Fig.~\ref{fig:tbot},
the interlayer hopping does harm to the superconductivity in our model with fixed $U$ and $J$.
The reason could be that $t_{\bot}$ prefer bonding and antibonding state, and large $t_{\bot}$ push these states away from fermi surface, then many-body effects
will be suppressed. In fact, the strength in $d_{z^2}$ can achieve $t_{\bot} / t > 5$, which is examined in Fig.~\ref{fig:tbot}.
We compare superfluid density $\rho_s$ with various $t_{\bot}$, and $T_{c}$ can be estimated  by
finding the crossing point of $\rho_s$ and $2T/ \pi$. One can see that large $t_{\bot}$
does not favor superconductivity.

However, in the simulations above, we have fixed the strength of interactions $J$ and $U$. In real materials, the model parameters could be more complicated.
The interlayer spin-exchange term $J$ is originated from $U$ and $t_{\bot}$,
and increasing $t_{\bot}$ will also enhance $J$. This is different from our model where the proportion of $U$ and $J$ is fixed.
In real LNO system, a medium $t_{\bot}$ should be beneficial to superconductivity, and further increasing $t_{\bot}$ is harmful to superconductivity.
A thoughtful investigation of this competition is interesting but beyond the scope of this paper.

\begin{figure}
    \centering
    \includegraphics[width=0.45\textwidth]{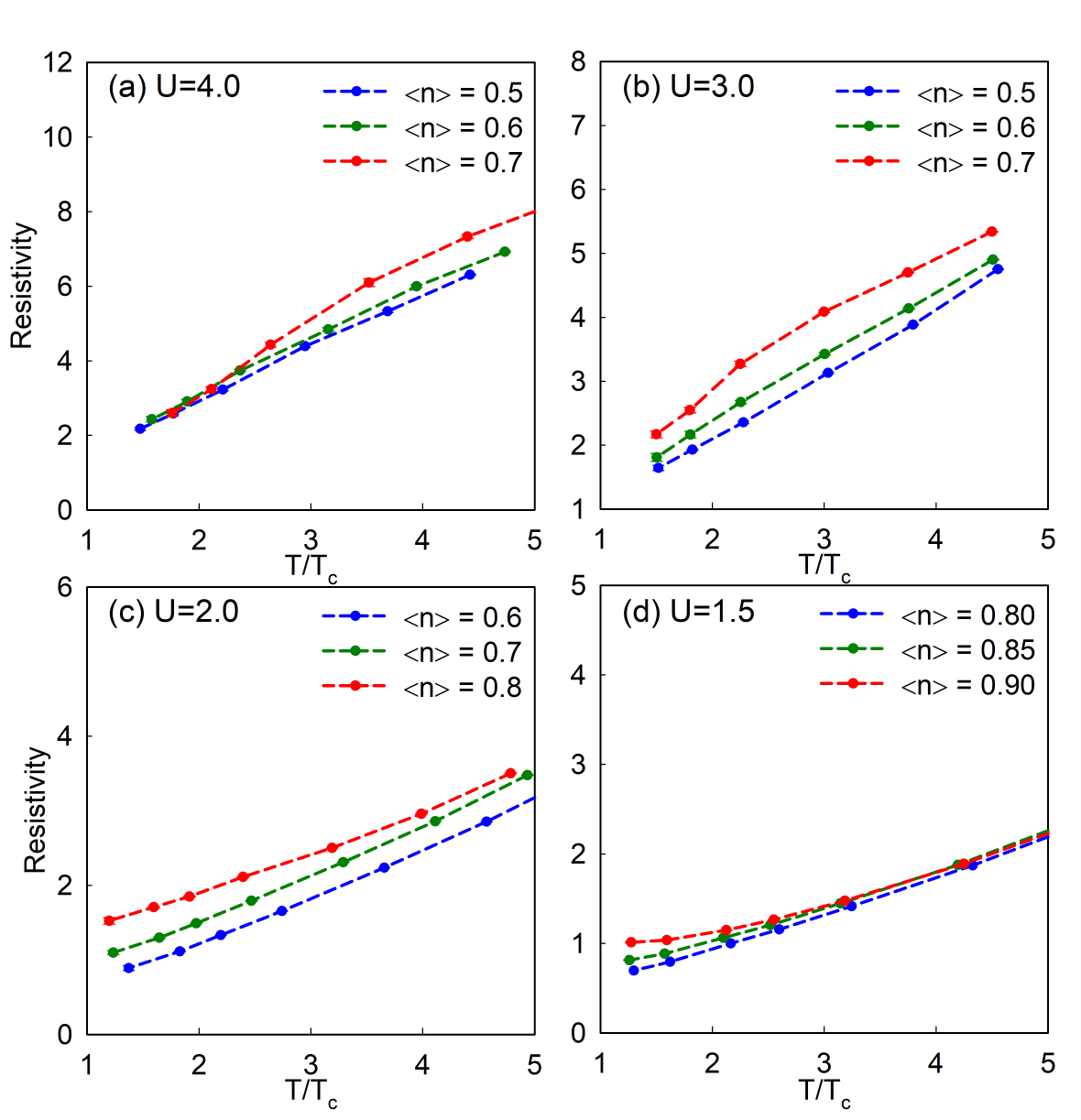}
    \caption{Resistivity at (a) $U=4.0$ and (b) $U=1.5$. Where the resistivity shows a linear dependence
    on $T$ at relative high temperatures.}
    \label{fig:resis}
\end{figure}

In the studies of high-$T_{c}$ superconductors, strange-metal behavior which develops in the normal state, is of great importance for theoretical understanding.
As we can calculate current-current correlations to determine the superfluid density, it is readily to extract the regular part of resistivity $R^{reg}$ by the relation
shown in Eq.~\ref{eq:10}.
In Fig.~\ref{fig:resis}, we show the resistivity at relative high temperature as the method we used is only compatible at normal state.
In this picture, we can see a linear relationship between temperature $T$ and resistivity, and this relation is more clear at lower density $\langle n \rangle$.
This may indicate a strange-metal behavior at relative high temperature.

\noindent
\underline{\it Conclusions}~~
We study a sign-problem-free bilayer Hubbard-like model by using DQMC algorithm, and this model is closely related with recently discovered bilayer nickelate superconductor $\mathrm{La}_{3}\mathrm{Ni}_{2}\mathrm{O}_{7}$.
Both the effective model for LNO and our model emphasize the significant role of the spin-exchange $J$ term in the formation of superconductivity.
Moreover, due to the special symmetry our model holds, we can get rid of sign problem in simulations with arbitrary electron filling,
and this permits us to conduct DQMC simulations at relatively low temperatures with sufficient accuracy.
Our unbiased numerical calculations show that the optimal doping is moving to $\langle n \rangle = 0.5$ as the interaction become stronger.
In fact, this reflects the complex dependence on $U$. It enhances superconductivity at high doping and
weaken it at low doping. And the interlayer hopping term also weaken the superconductivity.
Furthermore, the regular part of resistivity is calculated, and the possible strange-metal behaviors of this model are discussed.
In summary, our work investigates the superconducting transition temperature $T_c$ in a bilayer Hubbard-like system
and the factors that impact this transition, enhances our understanding of high-temperature superconductivity.

\bibliography{ref}

\end{document}